\documentclass[12pt]{article}

\newcommand{\be}{\begin{equation}}
\newcommand{\ee}{\end{equation}}
\newcommand{\bi}[1]{\vspace{-3mm} \bibitem{#1}}
\usepackage{epsfig,amsmath,amssymb,graphics,graphicx}

\begin{document}

\begin{center}
Journal of Mathematical Physics 50 (2009) 122703
\end{center}

\begin{center}

{\Large \bf  Discrete Map with Memory from Fractional Differential Equation 
of Arbitrary Positive Order}
\vskip 5 mm

{\large \bf Vasily E. Tarasov}\\

\vskip 3mm

{\it Skobeltsyn Institute of Nuclear Physics, \\
Moscow State University, Moscow 119991, Russia } \\
{E-mail: tarasov@theory.sinp.msu.ru}

\vskip 11 mm

\begin{abstract}
Derivatives of fractional order with respect to time 
describe long-term memory effects. 
Using nonlinear differential equation 
with Caputo fractional derivative of arbitrary order $\alpha>0$,
we obtain discrete maps with power-law memory.
These maps are generalizations of well-known universal map.
The memory in these maps means that their present state is determined 
by all past states with power-law forms of weights. 
Discrete map equations are obtained by using
the equivalence of the Cauchy-type problem 
for fractional differential equation and
the nonlinear Volterra integral equation of the second kind.
\end{abstract}

\end{center}

\section{Introduction}

Discrete maps are used for the study of evolution problems, 
possibly as a substitute of differential equations \cite{Chirikov,SUZ,Zaslavsky1}. 
They lead to a much simpler formalism, which is particularly useful 
in numerical simulations.
The universal discrete map is one of the most widely studied maps.
It is a very important step in understanding the qualitative
behavior of a wide class of systems described by differential equations.
The derivatives of noninteger orders \cite{KST,Podlubny,SKM} 
are a natural generalization of the ordinary differentiation of integer order. 
Fractional differentiation with respect to time is
characterized by power-law memory effects. 
The discrete maps with memory are considered in Refs. 
\cite{Ful,Fick1,Fick2,Giona,Gallas,Stan,JPA2008}. 
It is important to connect fractional differential equations 
and discrete maps with memory.
In Ref. \cite{JPA2008}, we prove that the discrete maps with memory 
can be derived from differential equations with fractional derivatives.
The fractional generalization of the universal map 
was derived \cite{JPA2008} from a differential equation 
with Riemann-Liouville fractional derivatives.
The Riemann-Liouville  derivative has some notable 
disadvantages in physical applications such as 
the hyper-singular improper integral, where the order 
of singularity is higher than the dimension, and 
nonzero of the fractional derivative of constants, 
which would entail that dissipation does not vanish for a 
system in equilibrium.  
The desire to formulate initial value problems for physical
systems leads to the use of Caputo fractional derivatives 
rather than Riemann-Liouville fractional derivatives.
In this paper, we obtain a discrete map with memory 
from differential equations with Caputo fractional derivative
of arbitrary order $\alpha>0$.
The universal map with power-law memory 
is obtained by using the equivalence
of the fractional differential equation and the Volterra integral equation.
We reduce the Cauchy-type problem for the differential equations 
with the Caputo fractional derivative  
to a nonlinear Volterra integral equation of the second kind.
The equivalence of this Cauchy-type problem 
and the correspondent Volterra equation 
was proved by Kilbas and Marzan in \cite{KM1,KM2}.  


In Section 2, differential equations with integer derivative and 
universal maps without memory are considered 
to fix notations and provide convenient references. 
In Section 3, fractional differential equations with Caputo derivative 
and correspondent discrete maps with memory are considered. 
A fractional generalization of the universal map is obtained from
kicked differential equations with the Caputo fractional 
derivative of arbitrary order $\alpha >0$.
Finally, a short conclusion is given in section 4.


\section{Universal map without memory}

In this section, differential equations with derivative 
of second order and the universal map without memory 
are considered to fix notations and provide convenient references. 

Let us consider the equation of motion, 
\be \label{eq1}
D^2_t x(t)+ K G[x(t)] \sum^{\infty}_{k=1} \delta \Bigl(\frac{t}{T}-k \Bigr)=0
\ee
in which perturbation is a periodic sequence of delta-function-type pulses (kicks)
following with period $T=2\pi / \nu$, $K$ is an amplitude of the pulses, 
$D^2_t=d^2/dt^2$, and $G[x]$ is some real-valued function. 
It is well known that this differential equation 
can be represented in the form of the discrete map, 
\be \label{UM}
x_{n+1}-x_n=p_{n+1}T , \quad p_{n+1}-p_n=-KT\, G[x_n] .
\ee
Equations (\ref{UM}) are called the universal map.
For details, see for example \cite{Chirikov,SUZ,Zaslavsky1}.

Traditional method of derivation of the universal map equations from
the differential equations is considered in Section 5.1 of Ref. \cite{SUZ}.
We use another method of derivation of these equations 
to fix notations and provide convenient references. 
It is easy to obtain the universal map by using the equivalence
of the differential equation and the Volterra integral equation. 
The Cauchy-type problem for the differential equations,  
\be \label{p}
D^1_t x(t) =p(t) ,
\ee
\be 
D^1_t p(t)=- 
K \, G[x(t)] \sum^{\infty}_{k=1} \delta \Bigl(\frac{t}{T}- k \Bigr) , 
\ee
with the initial conditions
\be \label{RL5}
x(0)=x_0, \quad p(0)=  p_0 
\ee
is equivalent to the universal map equations of the form 
\be \label{E6i}
x_{n+1}=x_0 + p_0 (n+1)T - K T^2 \sum^{n}_{k=1} G[x_k] \, (n+1-k) ,
\ee
\be \label{E7i}
p_{n+1}= p_0 - K T \sum^{n}_{k=1} G[x_k] .
\ee
To prove this statement we
consider the nonlinear differential equation (\ref{eq1}) 
on a finite interval $[0,t_f]$ of the real axis, 
with the initial conditions (\ref{RL5}).
The Cauchy-type problem of the form (\ref{eq1}) and (\ref{RL5}) 
is equivalent to the Volterra integral equation,  
\be \label{E3i}
x(t)=x_0+p_0 t - K \sum^{\infty}_{k=1} 
\int^t_{0} \, d\tau \, G[x(\tau)] \delta \Bigl(\frac{\tau}{T}-k \Bigr)  \, (t-\tau) .
\ee
For $nT<t<(n+1)T$, we obtain 
\be \label{E4i}
x(t)=x_0 + p_0 t - K T \sum^{n}_{k=1} G[x(kT)] \, (t-kT) .
\ee
Equations (\ref{E4i}) and (\ref{p}) give
\be \label{E5i}
p(t)= p_0 - K T \sum^{n}_{k=1} G[x(kT)] .
\ee
The solution of the left side of the $(n+1)$th kick
\be \label{not1}
x_{n+1}=x(t_{n+1}-0)=\lim_{\varepsilon \rightarrow 0+} x(T(n+1)-\varepsilon), 
\ee
\be \label{not2}
p_{n+1}=p(t_{n+1}-0)=\lim_{\varepsilon \rightarrow 0+} p(T(n+1)-\varepsilon), 
\ee
where $t_{n+1}=(n+1)T$ gives the map equations (\ref{E6i}) and (\ref{E7i}).  
This ends the proof.
Note that equations (\ref{E6i}) and (\ref{E7i}) 
can be rewritten in the form (\ref{UM}).   
Using equations (\ref{E6i}) and (\ref{E7i}), 
the differences $x_{n+1}-x_n$ and $p_{n+1}-p_n$ give 
equations (\ref{UM}) of the universal map.

We note that equations (\ref{UM}) with $G[x]=-x$ give the Anosov-type system
\be
x_{n+1}-x_n=p_{n+1}T , \quad p_{n+1}-p_n=KT x_n .
\ee
If $G[x]=\sin (x)$, then equations (\ref{UM}) are
\be
x_{n+1}-x_n=p_{n+1}T , \quad p_{n+1}-p_n=-KT \, \sin(x_n) .
\ee
This map is known as the standard or Chirikov map \cite{Chirikov}. 

\section{Fractional equation and universal map with memory}

In Ref. \cite{JPA2008} we consider nonlinear differential equations
with Riemann-Liouville fractional derivatives.
The discrete maps with memory  are obtained from these equations. 
The Riemann-Liouville fractional derivative has some notable 
disadvantages in physical applications such as 
the hypersingular improper integral, where the order 
of singularity is higher than the dimension, and 
nonzero of the fractional derivative of constants, 
which would entail that dissipation 
does not vanish for a system in equilibrium.  
The desire to formulate initial value problems for physical
systems leads to the use of Caputo fractional derivatives \cite{KST,Podlubny}
rather than Riemann-Liouville fractional derivative.


The left-sided Caputo fractional derivative \cite{Caputo,Caputo2,GM,KST}
of order $\alpha >0$ is defined by
\be \label{Caputo}
\,  _0^CD^{\alpha}_t f(t)=
\frac{1}{\Gamma(m-\alpha)} \int^t_0 
\frac{ d\tau \, D^m_{\tau}f(\tau)}{(t-\tau)^{\alpha-m+1}} =
\, _0I^{m-\alpha}_t D^m_t f(t) ,
\ee
where $m-1 < \alpha <m$ and $_0I^{\alpha}_t$ is 
the left-sided Riemann-Liouville fractional integral 
of order $\alpha >0$, that is defined by
\be
_0I^{\alpha}_t f(t)=\frac{1}{\Gamma(\alpha)} 
\int^t_0 \frac{f(\tau) d \tau}{(t-\tau)^{1-\alpha}} , \quad (t>0).
\ee
The Caputo fractional derivative first computes an ordinary
derivative followed by a fractional integral to achieve the
desire order of fractional derivative.
The Riemann-Liouville fractional derivative $\, _0D^{\alpha}_t $
is computed in the reverse order. 
Integration by part of (\ref{Caputo}) gives 
\be \label{C-RL}
\, _0D^{\alpha}_t x(t)= \, _0^CD^{\alpha}_t x(t)+
\sum^{m-1}_{k=0} \frac{t^{k-\alpha}}{\Gamma(k-\alpha+1)} x^{(k)}(0) .
\ee 
The second term in Eq. (\ref{C-RL}) regularizes 
the Caputo fractional derivative to avoid the potentially divergence 
from singular integration at $t=0$. In addition, the 
Caputo fractional differentiation of a constant results in zero
$ _0^CD^{\alpha}_t C=0$. 
The Riemann-Liouville fractional derivative of a 
constant need not be zero \cite{KST}.

If the Caputo fractional derivative is used instead of the 
Riemann-Liouville fractional derivative, 
then the initial conditions for fractional dynamical systems 
are the same as those for the usual dynamical systems.  
The Caputo fractional derivatives can be more 
applicable to dynamical systems than the Riemann-Liouville derivatives. 
Note that the Caputo fractional derivatives can be used to formulate
a self-consisted fractional vector calculus \cite{FVC}.


We consider the nonlinear differential equation of order $\alpha$, 
where $0 \le m-1 < \alpha \le m$,
\be \label{E1}
\,  _0^CD^{\alpha}_t x(t) = G[t,x(t)] , \quad (0 \le t \le t_f) ,
\ee
involving the Caputo fractional derivative $ _0^CD^{\alpha}_t$ 
on a finite interval $[0,t_f]$ of the real axis, 
with the initial conditions
\be \label{E2}
(D^k_t x)(0)=c_k , \quad k=0,...,m-1. 
\ee
Kilbas and Marzan \cite{KM1,KM2} proved the equivalence
of the Cauchy-type problem of the form (\ref{E1}), (\ref{E2}) and 
the Volterra integral equation of second kind 
\be \label{E3}
x(t)= \sum^{m-1}_{k=0} \frac{c_k}{k!} t^k + 
\frac{1}{\Gamma (\alpha)} \int^t_{0} d \tau \,
G[\tau,x(\tau)] \, (t-\tau)^{\alpha-1}
\ee
in the space $C^{m-1}[0,t_f]$.

The basic theorem regarding the nonlinear
differential equation involving 
the Caputo fractional derivative states that 
the Cauchy-type problem (\ref{E1}), (\ref{E2}) and 
the nonlinear Volterra integral equation (\ref{E3}) 
are equivalent in the sense that, if  $x(t) \in C [0,t_f]$
satisfies one of these relations, then it also satisfies the other. 
In \cite{KM1,KM2} (see also \cite{KST}, Theorem 3.24.) 
this theorem is proved by assuming that a function $G[t,x]$ 
for any $x \in W \subset \mathbb{R}$
belong to $C_{\gamma} (0,t_f)$ with $0 \le \gamma <1$, $\gamma <\alpha$.
Here $C_{\gamma} (0,t_f)$ is the weighted space 
of functions $f[t]$ given on $(0,t_f]$, such that  
$t^{\gamma} f[t] \in C(0,t_f)$. 


Let us consider a generalization of 
equation (\ref{eq1}) in the form of
the fractional differential equation, 
\be \label{E4}
_0^CD^{\alpha}_t x(t)+ 
K \, G[x(t)] \sum^{\infty}_{k=1} \delta \Bigl(\frac{t}{T}- k \Bigr)=0, 
\quad (m-1 <\alpha < m) ,
\ee
where $ _0^CD^{\alpha}_t$ is the Caputo fractional derivative, 
with the initial conditions 
\be \label{E5}
D^s_t x (0)=x^{(s)}_0  \quad (s=0,1,...,m-1).
\ee
Using $x^{(s)}(t)=D^s_t x(t)$, $s=0,1,...,m-1$, equation (\ref{E4}) 
can be rewritten in the Hamilton form. \\

\vskip 3mm
\noindent
{\large \bf Theorem.}
{\it The Cauchy-type problem for the fractional differential equations 
\be \label{mom}
D^1_t x^{(s)}(t) =x^{(s+1)}(t) , \quad (s=0,1,...,m-2)
\ee
\be 
_0^CD^{\alpha-m+1}_t x^{(m-1)}(t)=- 
K \, G[x(t)] \sum^{\infty}_{k=1} \delta \Bigl(\frac{t}{T}- k \Bigr) , 
\quad (m-1 <\alpha < m) ,
\ee
with the initial conditions
\be 
x^{(s)}(0)=x^{(s)}_0 , \quad (s=0,1,...,m-1)
\ee
is equivalent to the discrete map equations, }
\be \label{E9}
x^{(s)}_{n+1}=  \sum^{m-s-1}_{k=0} \frac{x^{(k+s)}_0}{k!} (n+1)^k T^k 
- \frac{KT^{\alpha-s}}{\Gamma(\alpha-s)} \sum^{n}_{k=1} 
\, (n+1-k)^{\alpha-1-s}  G[x_k] ,
\ee

\vskip 3mm
{\bf Proof.}
Using the Kilbas-Marzan result for equation (\ref{E1}) with the function
\[ G[t,x(t)]= - K G[x(t)] \sum^{\infty}_{k=1} \delta \Bigl(\frac{t}{T}-k \Bigr) , \]
we obtain that the Cauchy-type problem (\ref{E4}) and (\ref{E5}) is equivalent to
the Volterra integral equation of second kind, 
\be \label{E6}
x(t) = \sum^{m-1}_{k=0} \frac{x^{(k)}_0}{k!} t^k 
 - \frac{K}{\Gamma(\alpha)} 
\sum^{\infty}_{k=1} \int^t_{0} d \tau \, (t-\tau)^{\alpha-1} \, G[x(\tau)] 
\, \delta \Bigl(\frac{\tau}{T}- k \Bigr),
\ee
in the space of continuously differentiable functions $x(t) \in C^{m-1}[0,t_f]$. 

If $nT <t< (n+1)T$, then equation (\ref{E6}) gives
\be \label{E7}
x(t)= \sum^{m-1}_{k=0} \frac{x^{(k)}_0}{k!} t^k 
- \frac{K T}{\Gamma(\alpha)} \sum^{m}_{k=1} 
\, (t-kT)^{\alpha-1} \, G[x(kT)] .
\ee
Using the variables (\ref{mom}), equation (\ref{E7}) gives
\be \label{E8}
x^{(s)}(t)=\sum^{m-1-s}_{k=0} \frac{x^{(k+s)}_0}{k!} t^k 
 - \frac{KT}{\Gamma(\alpha-s)} 
\sum^{n}_{k=1} (t-kT)^{\alpha-1-s} \, G[x(kT)] ,
\ee
where $s=0,1,...,m-1$, $nT < t < (n+1)T$, $m-1<\alpha<m$
and we use $\Gamma(z)=(z-1) \Gamma(z-1)$.
The solution of the left side of the $(n+1)$th kick 
(\ref{not1}) and (\ref{not2}) can be represented by equations (\ref{E9}),
where we use the condition of continuity $x^{s}(t_n+0)=x^s(t_n-0)$, 
$s=0,1,...,m-2$. 

This ends the proof. $\ \ \ \Box$ \\

Equations (\ref{E9}) define
a generalization of the universal map.
This map is derived from a fractional differential equation 
with Caputo derivatives without any approximations.
The main property of the suggested map 
is a long-term memory that means that their present state 
depends on all past states with a power-law form of weights. 

If $G[x]=\sin(x)$, then equations (\ref{E9}) define 
a generalization of standard map. 
For $G[x]=-x$, we have Anosov-type system with memory.

In the case of $1<\alpha<2$, $m=2$, 
we have the following universal map with memory:
\be
x_{n+1}=x_0+p_0(n+1)T - 
\frac{KT^{\alpha}}{\Gamma(\alpha)} \sum^{n}_{k=1} \, (n+1-k)^{\alpha-1}  G[x_k] ,
\ee
\be 
p_{n+1}=p_0 - 
\frac{KT^{\alpha-1}}{\Gamma(\alpha-1)} \sum^{n}_{k=1} \, (n+1-k)^{\alpha-2}  G[x_k] .
\ee
where $x_n=x^{(0)}_n$ and $p_n=x^{(1)}_n$. 
If $\alpha=m=2$, then equations (\ref{E9}) give 
the universal map of the form (\ref{E6i}) and (\ref{E7i})
that is equivalent to equations (\ref{UM}). 
As a result, the usual universal map is a 
special case of this universal map with memory.

\section{Conclusion}

Equations for discrete maps with memory are suggested.
The maps with power-law memory
describe fractional dynamics of complex physical systems.  
The suggested map with memory is a generalization of well-known universal map.
These maps are equivalent to the
correspondent fractional kicked differential equations.
To derive the map equations an approximation 
for fractional derivatives is not used.
We obtain a discrete map with memory
from fractional differential equation
by using the equivalence of the Cauchy-type problem and 
the nonlinear Volterra integral equation of the second kind.

Fractional differentiation with respect to time is
characterized by power-law memory effects that 
correspond to intrinsic dissipative processes 
in the physical systems. 
Therefore, the universal maps with memory 
have regular and strange attractors for 
some values of parameters $K$ and $\alpha$.
The suggested universal maps with memory demonstrate 
a chaotic behavior with a new type of attractors.
Numerical simulations of the universal map with memory 
prove that the nonlinear dynamical systems, 
which are described by the equations with fractional derivatives, 
exhibit a new type of chaotic motion. 
For some regions of parameters $K$ and $\alpha$ 
these universal maps with memory demonstrate 
a new type of regular and strange attractors.
The universal maps with power-law memory can be used  
to describe properties of regular and strange attractors
of the fractional differential equations with kicks.



\end{document}